\documentclass[fleqn,usenatbib]{mnras}

\usepackage{newtxtext,newtxmath}

\usepackage[T1]{fontenc}
\usepackage{ae,aecompl}

\usepackage{graphicx}	
\usepackage{amsmath}	

\newcommand{\zph}{$z_{\rm phot}$}

\newcommand{\Msun}{M$_{\odot}$}

\title[ASKAP discovery of ORC J0102--2450]{Discovery of a new extragalactic circular radio source with ASKAP: ORC~J0102--2450}

\author[Koribalski et al.]{B\"arbel S. Koribalski,$^{1,2}$\thanks{E-mail: Baerbel.Koribalski@csiro.au}
Ray P. Norris,$^{2,1}$ 
Heinz Andernach,$^{3}$
Lawrence Rudnick,$^{4}$
\newauthor
Stanislav Shabala,$^{5}$
Miroslav Filipovi\'c,$^{2}$ 
and Emil Lenc$^{1}$ \\
$^{1}$Australia Telescope National Facility, CSIRO Astronomy and Space Science, P.O. Box 76, Epping, NSW 1710, Australia \\
$^{2}$School of Science, Western Sydney University, Locked Bag 1797, Penrith, NSW 2751, Australia \\
$^{3}$Departmento de Astronom\'ia, Universidad de Guanajuato, Callejón de Jalisco s/n,
Guanajuato, C.P. 36023, GTO, Mexico \\
$^{4}$Minnesota Institute for Astrophysics, University of Minnesota, 116 Church St. SE, Minneapolis, MN 55455, USA \\
$^{5}$School of Natural Sciences, University of Tasmania, Private Bag 37, Hobart 7001, Australia
}

\date{Accepted XXX. Received YYY; in original form ZZZ}

\pubyear{2021}

\begin{document}
\label{firstpage}
\pagerange{\pageref{firstpage}--\pageref{lastpage}}
\maketitle

\begin{abstract}
We present the discovery of another Odd Radio Circle (ORC) with the Australian Square Kilometre Array Pathfinder (ASKAP) at 944 MHz. The observed radio ring, ORC~J0102--2450, has a diameter of $\sim$70 arcsec or 300 kpc, if associated with the central elliptical galaxy DES~J010224.33--245039.5 ($z \sim 0.27$). Considering the overall radio morphology (circular ring and core) and lack of ring emission at non-radio wavelengths, we investigate if ORC~J0102--2450 could be the relic lobe of a giant radio galaxy seen end-on or the result of a giant blast wave. We also explore possible interaction scenarios, for example, with the companion galaxy, DES~J010226.15--245104.9, located in or projected onto the south-eastern part of the ring. We encourage the search for further ORCs in radio surveys to study their properties and origin.

\end{abstract}

\begin{keywords}
radio continuum: galaxies, ISM -- instrumentation: radio interferometers
\end{keywords}



\section{Introduction} 
\label{sec:intro}

Odd radio circles (ORCs), first discovered by Norris et al. (2021a) in ASKAP radio continuum data from the `Evolutionary Map of the Universe' (EMU) Pilot Survey (800 -- 1088 MHz; rms $\sim$30 $\mu$Jy\,beam$^{-1}$), resemble rings or edge-brightened disks of radio emission that, so far, remain undetected at non-radio wavelengths. In their paper, Norris et al. present four ORCs (each $\sim$60 arcsec in diameter), three of which were detected with ASKAP, including a pair of ORCs. The fourth ORC, which is notable for its central radio source, was discovered in 325 MHz radio data from the Giant Meterwave Radio Telescope (GMRT). Most notably, the two single ORCs each have an elliptical galaxy in their geometrical ring centre. 

Odd Radio Circles -- at first glance -- look like supernova remnants (SNRs), so could they be formed by a giant blast wave from a transient event (e.g., a merging binary super-massive black hole (SMBH), a hyper-nova, or a $\gamma$-ray burst) in the central elliptical galaxy many millions of years ago\,? In Norris et al. (2021a) a wide range of possible formation scenarios are discussed, which are evaluated and expanded with each new ORC discovery.

In this paper we focus on ORC J0102--2450, newly discovered during the search for ORCs and other extended radio sources in a deep $\sim$40 deg$^2$ ASKAP field centred near the starburst galaxy NGC~253. A summary of the ASKAP multi-epoch observations is given in Section~2, followed by our analysis of the ORC J0102--2450 properties in Section~3. Possible formation scenarios are discussed in Section~4, and our conclusions are given in Section~5. 

\begin{figure*} 
\centering
 \includegraphics[width=14cm]{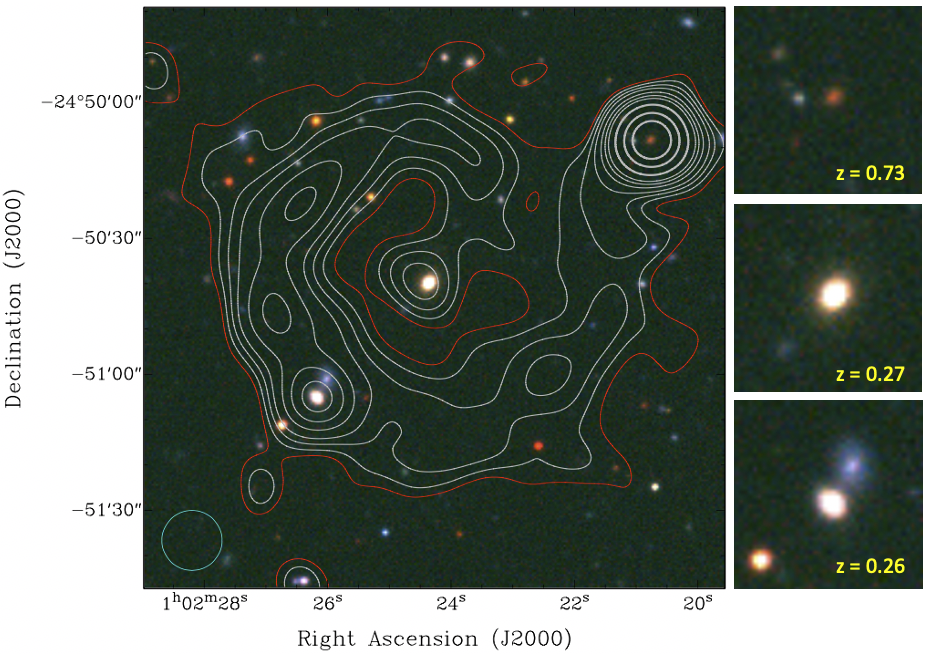}
\caption{ORC J0102--2450. --- ASKAP radio continuum contours overlaid onto an optical RGB colour image  created from the Dark Energy Survey (DES) $zrg$-bands. The ASKAP radio contour levels are 0.045 (dark red), 0.065, 0.09, 0.12, 0.17, 0.22, 0.27, 0.4, 0.6 and 0.8 mJy\,beam$^{-1}$; the resolution is 13$''$ and the rms is $\sim$15$\mu$Jy\,beam$^{-1}$. DES image cutouts of the three radio-detected galaxies and their average photometric redshifts are shown on the right side (top: background galaxy, middle: central galaxy, bottom: south-eastern galaxy).}
\label{fig:orc-fig1}
\end{figure*}

\section{ASKAP Observations and Data Processing}
\label{sec:obs}

ASKAP is a new radio interferometer consisting of 36 $\times$ 12-m antennas, each equipped with a wide-field Phased Array Feed (PAF), baselines out to 6.4~km, operating at frequencies from 700 MHz to 1.8 GHz (Johnston et al. 2008). The currently available bandwidth of 288~MHz is divided into $288 \times 1$ MHz coarse channels; for a comprehensive system overview see Hotan et al. (2021). We obtained nine fully calibrated ASKAP radio continuum images ($\sim$10h integration time each) from the CSIRO ASKAP Science Data Archive (CASDA)\footnote{CASDA: \url{dap.csiro.au}}. The field centres are close to the nearby starburst galaxy NGC~253 (HIPASS J0047--25; PKS J0047--2517), which has a total integrated flux density of 6.2 Jy at 1.4 GHz and 8.1 Jy at 960 MHz (K\"uhr et al. 1981). The brightness and large extent of the NGC~253 star-forming disc causes minor artefacts over most of the field at a flux level of $\lesssim$1\%. The nine ASKAP observations were conducted between Aug 2019 and Dec 2020 with the band centred at 943.5 MHz. ASKAP PAFs were used to form 36 beams arranged in a $6 \times 6$ closepack36 footprint (Hotan et al. 2021), each delivering a $\sim$30 deg$^2$ field of view out to the half power point. The data processing was done in the ASKAPsoft pipeline (Whiting et al. 2017). We combined eight of the nine ASKAP radio continuum images (leaving out the last, rather short observation) after convolving each to a common 13\arcsec\ resolution, achieving an rms sensitivity of $\sim$15 $\mu$Jy\,beam$^{-1}$ near ORC~J0102--2450. 

\vspace{-0.5cm}

\section{ORC J0102--2450}
Figures~1 \& 2 show the deep ASKAP radio continuum images of ORC~J0102--2450 together with DES optical (Abbott et al. 2018) and WISE infrared (Wright et al. 2010) images. The ORC properties are listed in Table~1 and compared to similar ORCs in Table~2. In the ring centre we find an elliptical galaxy (2MASS~J01022435--2450396, DES~J010224.33--245039.5) which has a photometric redshift of $z \approx 0.27$ (Bilicki et al. 2016, Zou et al. 2019). Anchored on this galaxy, we measure a ring diameter of $\sim$70 arcsec which matches the radio emission peaks along its circumference. If associated, this suggests an ORC diameter of $\sim$300 kpc. The eastern side of the ring is significantly brighter and narrower than the western side, resulting in a {\bf C}-like structure. We measure an average ring width of $\sim$25 arcsec from the ridge line to the 50\% intensity. The ring width on the eastern side is $\sim$19 arcsec or $\sim$14 arcsec deconvolved. A polar projection of the ORC is shown in Fig.~3. In addition to the central radio core/jet, we detect diffuse radio emission inside and outside of the ring, extending out to a diameter of $\sim$120 arcsec (500 kpc). Although this emission is faint, it may account for a significant fraction of the total ORC flux.

No optical or infrared counterparts to the radio ring have been detected, apart from a luminous infrared galaxy (LIRG) in the south-eastern section, which is discussed below. The weak central radio source is located at $\alpha,\delta$(J2000) = $01^{\rm h}\,02^{\rm m}\,24.5^{\rm s}$, $-24^{\circ}\,50'\,38''$ ($\sim$0.1 mJy), coincident with a bright elliptical galaxy (see Table~1). The classification of this galaxy as an elliptical, based on the WISE colour-colour diagram (Jarrett et al. 2017), is likely. From the 2MASS $K_{\rm s}$-band luminosity of the central galaxy in ORC J0102--2450 we estimate a black hole mass of $\sim$7.5 $\times 10^8$~\Msun, following Graham (2007). The total integrated flux density of ORC~J0102--2450 as measured in our ASKAP 944 MHz data is $\sim$3.9 mJy. It is clearly detected in Stokes $I$ in each of the nine epochs; no polarisation has been detected.

\begin{table}
\centering
\caption{ASKAP 944 MHz properties of ORC J0102--2450 and optical/infrared properties of the galaxies in the ring centre and the south-eastern (SE) part of the ring.}
\begin{tabular}{ll}
\hline
  source name & ORC J0102--2450 \\
\hline
  radio core position & $01^{\rm h}\,02^{\rm m}\,24.5^{\rm s}$, $-24^{\circ}\,50'\,38.1''$ \\
  radio ring diameter & 70 arcsec (300 kpc) \\
  average ring width & $\sim$25 arcsec \\
  total radio flux & $\sim$3.9 mJy \\
  total radio luminosity & $\sim$5.5 $\times\ 10^{23}$ W\,Hz$^{-1}$ \\
  core radio flux  & $\sim$0.1 mJy (2.6\% of total) \\
  core radio luminosity & $\sim$1.4 $\times\ 10^{22}$ W\,Hz$^{-1}$ \\
\hline
  core galaxy names 
  & DES J010224.33--245039.5 \\
  ~~~~(ring centre) 
  & 2MASS J01022435--2450396 \\ 
  & WISEA J010224.35--245039.6 \\
  & CWISE J010224.33--245039.6 \\
  photometric redshift & 0.2753$^{\rm a}$, 0.2696$^{\rm b}$ \\
  optical galaxy size$^{\rm c}$ & $0.12' \times 0.07'$, $PA \sim 147^{\circ}$ \\
  & (30 kpc $\times$ 18 kpc)\\
  galaxy stellar mass$^{\rm b}$ & $\sim$10$^{11}$ M$_{\odot}$ \\
  2MASS $K_{\rm s}$ [mag] 
           & $15.1 \pm 0.2$ 
           ($4''$ aperture) \\
  $->$ black hole mass & $\sim$7.5 $\times 10^8$ M$_{\odot}$ \\
  WISE$^{\rm d}$ W1,2,3 [mag]
           & 15.39, 15.19, $>$12.36 (profile fit) \\
  DES$^{\rm c}$ $grizY$ [mag] & 20.43, 18.97, 18.48, 18.18, 18.03 \\
  log luminosities$^{\rm b}$ & 7.7 (NUV), 9.8 ($R$), 9.3 ($K$) \\
\hline
  SE galaxy names & DES J010226.15--245104.9 \\
  & WISEA J010226.16--245104.9 \\
  & CWISE J010226.15--245104.9 \\
  photometric redshift & 0.2835$^{\rm a}$, 0.2463$^{\rm b}$ \\
  optical galaxy size & $0.12' \times 0.08'$, $PA \sim 43^{\circ}$ \\
  galaxy stellar mass$^{\rm b}$ & $\sim$1.4 $\times 10^{10}$ M$_{\odot}$ \\
  WISE$^{\rm d}$ W1,2,3 [mag] & 15.60, 14.86, 11.16 (profile fit) \\
  DES$^{\rm c}$ $grizY$ [mag] & 20.48, 19.48, 18.99, 18.75, 18.60 \\
  log luminosities$^{\rm b}$ & 10.3 (NUV), 10.2 ($R$), 9.2 ($K$) \\
\hline
\end{tabular}
{\flushleft {\small (a) Bilicki et al. (2016), (b) Zou et al. (2019), see CDS J/ApJS/242/8/, (c) Abbott et al. (2018), (d) Wright et al. (2010), Cutri et al. (2013). }}
\label{tab:radio-orc}
\end{table}

A prominent radio peak within the south-eastern (SE) part of the ring ($\sim$0.25~mJy\,beam$^{-1}$) corresponds to the galaxy WISEA~J010226.15--245104.9 (DES~J010226.15--245104.9). It has a similar redshift to the central galaxy (see Table~1), suggesting they form a physical galaxy pair and are likely interacting. The WISE colours (see Fig.~2) suggest it is a star-forming galaxy or LIRG (Jarrett et al. 2017). The DES $g-r$ colour of 1.5 is very red for a star-forming galaxy, so this may be a dusty LIRG. Andernach (priv. comm.) finds that $g-r$ = 1.5 is typical for the hosts of radio galaxies at redshift $\sim$0.3. We note that the ORC emission is weakest opposite the SE galaxy (see Section~4.3). The relatively bright ($\sim$1~mJy) radio source north-west of ORC~J0102--2450 corresponds to the distant (red) elliptical galaxy WISEA~J010220.79--245008.6 (DES~J010220.74--245008.0) at \zph\ = 0.731 (Zou et al. 2019), and is unrelated to the ORC.

ORC J0102--2450 has a possible unresolved low-frequency counterpart, GLEAMSGP J010224--245024 (Franzen et al. 2021), but data at higher resolution and sensitivity are needed to confirm this. Using the fitted GLEAMSGP 200 MHz (resolution $\sim$2$'$) and the ASKAP 944 MHz flux densities for the same area ($17 \pm 1$ mJy and $\sim$5 mJy, respectively), i.e. the ORC and neighboring $\sim$1~mJy point source, we derive a 200--944 MHz spectral index of $\alpha = -0.8 \pm 0.2$, assuming $S \propto \nu^{\alpha}$ where $S$ is the flux density at frequency $\nu$.

\begin{figure*} 
\centering
  \includegraphics[width=17cm]{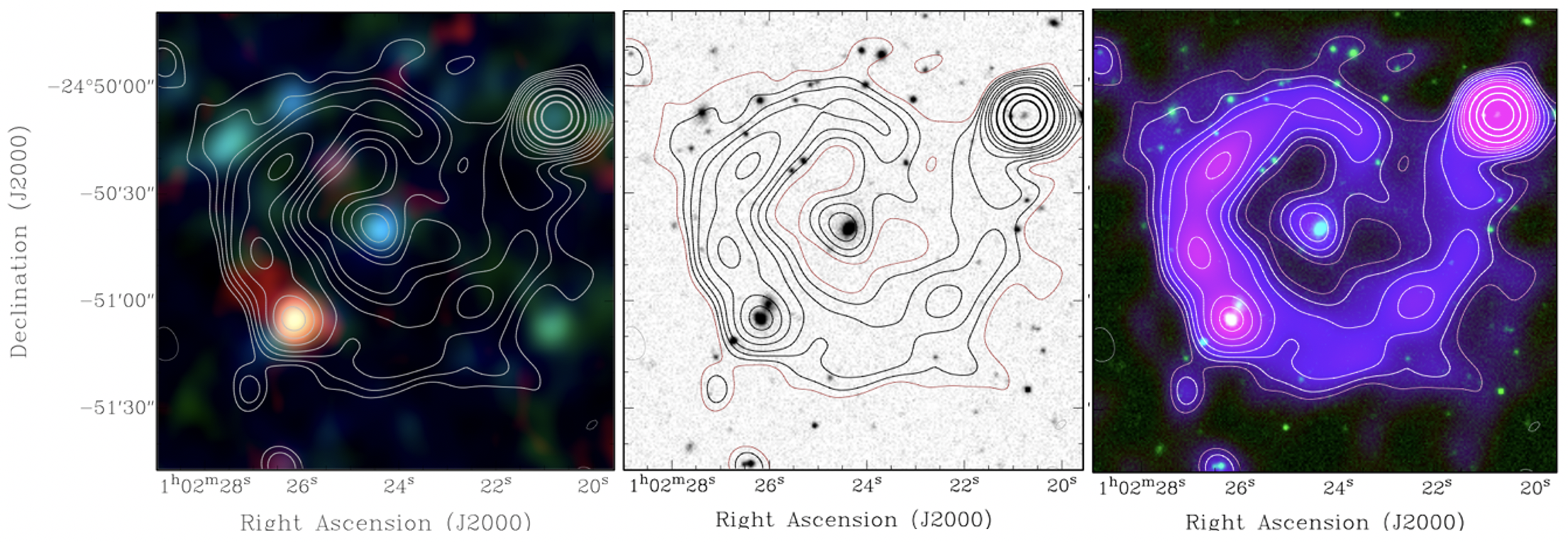}
\caption{ASKAP radio continuum contours of ORC J0102--2450 overlaid onto a WISE RGB colour image (red: 12$\mu$m (W3), green: 4.6$\mu$m (W2), and 3.4$\mu$m (W1)} -- at 7 arcsec resolution; {\bf left}), a DES $r$-band grey-scale image ({\bf middle}), and an ASKAP/DES combined RGB colour image, where the ASKAP image is shown in a combination of blue and red, while the DES $r$-band image is coloured green ({\bf right}). The ASKAP radio contour levels are --0.045, 0.045 (3$\sigma$), 0.065, 0.09, 0.12, 0.17, 0.22, 0.27, 0.4, 0.6 and 0.8 mJy\,beam$^{-1}$; the resolution is 13$''$ and the rms is $\sim$15$\mu$Jy\,beam$^{-1}$.
\label{fig:orc-fig2}
\end{figure*}

\begin{figure} 
\centering
 \includegraphics[width=6.5cm]{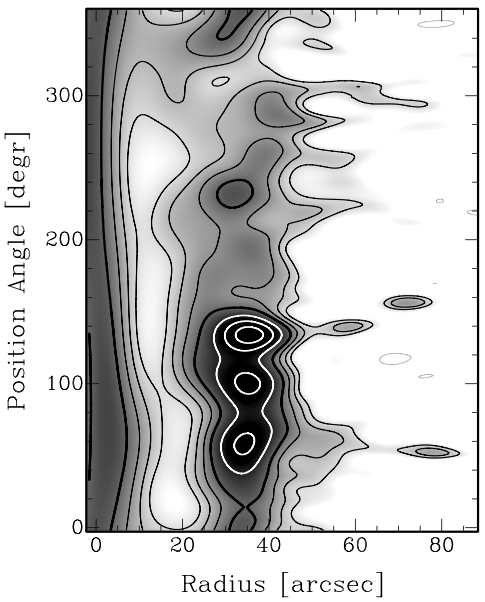}
\caption{ORC J0102--2450. --- Polar projection of the ASKAP radio continuum emission centred on the radio core; contour levels are the same as in Fig.~2. The position angle starts in the North (0 deg) proceeding  counter-clockwise. The radio point source north-west of the ORC was subtracted from the data prior to creating this image.}
\label{fig:orc-fig3}
\end{figure}

\vspace{-0.5cm}

\section{Discussion}
Here we discuss possible explanations for the origin of ORCs. Of the four ORCs discussed by Norris et al. (2021a), two --- ORCs~1 \& 4 --- have clear similarities to our discovery. See Table~2 for a brief summary of their properties. Most notably, each of the three ORCs has an elliptical galaxy in their respective ring centre, which is unlikely a coincidence. The probability of finding a radio source randomly coincident with the central 10$''$ of the ring is only $5 \times 10^{-3}$, based on the EMU source density of 814 deg$^{-2}$ for a 5$\sigma$ threshold (Norris et al., in prep.). While a class of three objects is very small, further candidate ORCs with similar features are under investigation. The double ORC feature (ORCs 2 \& 3) is distinctly different with neither having an associated host galaxy nor a central radio source and is not further considered here.

The three ORCs have similar sizes, morphologies, brightness, redshifts, spectral indices and central elliptical host galaxies, as summarised in Table~2. Taking into account recent MeerKAT 20-cm observations of ORC~1 (Norris et al., priv. comm.), all three have central radio sources, in positional agreement with their proposed elliptical host galaxies. In the following we focus on possible ORC formation scenarios where the observed radio rings are primarily associated with the central elliptical galaxies. Could ORCs be the projected view of a giant radio galaxy lobe seen end-on\,? Or are we seeing a new class of radio source\,? The similarity in morphology to Galactic SNRs suggests an energetic transient event may be responsible for a spherical shock wave of radio emission. Or could ORCs be the result of complex interactions, where the bent radio tails of one or two galaxies form a ring-structure\,?

\begin{table*}
\centering
\caption{Properties of three ORCs with central galaxies.}
\begin{tabular}{lcccccccc}
\hline
 \multicolumn{2}{c}{source name} & discovery & central & galaxy & \multicolumn{2}{c}{ring diameter} & spectral & Ref. \\
 & & telescope & host galaxy & redshift & [arcsec] & [kpc] & index \\
 \hline
 ORC J2103--6200 & (ORC~1) & ASKAP & WISE J210258.15--620014.4 & 0.55 & 80 & 510 & $-1.17 \pm 0.04$  & Norris et al. 2021a \\
 ORC J1555+2726 & (ORC~4)  & GMRT  & WISE J155524.65+272633.7  & 0.39  & 70 & 370 & $-0.92 \pm 0.18$ & Norris et al. 2021a \\
 ORC J0102--2450 & (ORC~5) & ASKAP & DES J010224.33--245039.5  & 0.27 & 70 & 300 & $-0.8 \pm 0.2$ & this paper \\
 \hline
\end{tabular}
\label{tab:ORC-comparison}
\end{table*}

\subsection{Giant radio lobes seen end-on} 

One possible explanation is that ORCs are relic radio galaxy lobes seen end-on, possibly in a rare transition state where the central jet has switched off but the shells of fading lobes are still expanding into the surrounding intergalactic medium (IGM). The circular shapes of ORCs and their central (host) galaxies require  alignment of the radio shells with the observer's line-of sight, suggesting they are unperturbed by the surrounding IGM. The ORC diameters of 300 -- 500 kpc suggest that we are dealing with giant radio galaxies (GRGs), the elliptical hosts galaxies of which lie in the respective ring centres. Given the ORC spectral indices (see Table~2), higher resolution observations, especially at lower frequencies, are expected to detect many more ORCs. Radio galaxies seen oriented in such a way must exist, but may not necessarily look like ORCs. An example of a relic giant radio galaxy with nearly straight lobes that fit this scenario is presented by Tamhane et al. (2015). Viewed end-on, it may resemble an ORC.

This scenario is supported by 2D hydrodynamical simulations (e.g., Br\"uggen \& Kaiser 2001) who investigate the expansion and shape of radio lobes after the SMBH-driven jet has turned off, finding large bubbles and toroidal shapes that may resemble ORCs when seen end-on. These appear to be short-lived unless re-energised (e.g., Subrahmanyan et al. 2008; Murgia et al. 2011; Tamhane et al. 2015), contributing to the rarity of ORCs. Nolting et al. (2019) in their simulations find "vortex rings" --- which resemble ORCs --- resulting from a shock by a head-on collision of a radio galaxy lobe with a stationary plane of hot intracluster gas. Detailed 3D simulations of the remnant radio galaxy scenarios are currently underway, and will be reported in a subsequent publication. 

\subsection{Radio remnant of a giant blast wave}
Could ORCs be the result of a giant blast wave in the central galaxy, producing a spherical shell of radio emission\,? Such shells would appear as edge-brightened discs, similar to supernova remnants or planetary nebulae. The radio emission is presumably synchrotron emission from electrons accelerated by a shock (e.g., Downes et al. 2002; Cao \& Wang 2012; Ricci et al. 2021). The edge-brightened emission is caused by the long path length through the limbs of the sphere, while the diffuse emission within the ring is caused by the front and back hemispheres of the shell. The thickness of the shell would determine the ratio of the ring emission brightness to the diffuse emission brightness. One way to produce such a spherical shock would be as the result of a binary SMBH merger in the central galaxy (e.g., Bode et al. 2012). In that case, we would expect to observe a largely tangential magnetic field in the ring, orthogonal to the velocity of the shock, similar to that of supernova remnants or cluster relics. Assuming the blastwave evolution is dominated by the Sedov-Taylor phase, for a fiducial ICM density of $10^{-3}$\,cm$^{-3}$ and a sound speed of 1000 km\,s$^{-1}$ we estimate a blastwave age of $t = 57/M$ Myr and energy $E = (6 \times 10^{53} / M^2)$ Joules, where $M$ is the current Mach number of the expanding shell. For transonic Mach numbers, this energy corresponds to a mass deficit of several million $M_\odot$, broadly consistent with a supermassive black hole merger scenario.

\subsection{Interaction scenarios}
Could the galaxy DES~J010226.15--245104.9 ($z \sim 0.26)$, located in the radio ring $\sim$150 kpc southeast of the ORC centre (SE galaxy in Table~2), possibly be the host of a radio galaxy with two narrow tails curving to form an incomplete ring\,? This scenario arises for ORC~J0102--2450 because of its {\bf C}-shaped structure, where the eastern, somewhat brighter side of the radio ring could be interpreted as a radio tail emerging northwards from the SE galaxy. The galaxy's radio flux density is $\sim$0.25 mJy. In this scenario, each of the narrow, curved radio tails would span $\sim$300 kpc. This may also be applicable to ORC~1 (Norris et al. 2021a) with DES~J210257.79--620045.9 (\zph\ = 0.23) in the southern part of the ring being the possible host galaxy. Narrow- and wide-angle tail radio galaxies are generally found in merging clusters where it is proposed that their motion through the intracluster medium (ICM) shapes their lobes (e.g., Smol\v{c}i\'c et al. 2007, and references therein). The ORCs do not appear to reside in clusters, and this scenario fails to explain their central galaxies. \\

Could interactions between the two galaxies and their radio lobes be responsible for the observed ORC shape\,? For spiral galaxies, interactions typically result in tidal tails, bridges and filaments, which can have a wide range of shapes and sizes (for a review see Koribalski 2019). A beautiful example of gravitational and ram pressure interaction between two double-lobe radio galaxies exist in the Abell 3785 cluster. The radio galaxy pair PKS~2130--538 (Jones \& McAdam 1992) with host galaxies PGC~66975/6 (both at $z \approx 0.077$), separated by $\sim$350 kpc, is also detected in the EMU Pilot Survey (Norris et al., in prep.). It seems unlikely that two interacting radio galaxies would be able to form the observed radio ring. 
Simulations of helical jets have succeeded in creating radio rings, like the doughnut shaped counter-minilobe of NGC~6109 which lies opposite a long (helical) radio tail (Rawes et al. 2018). A helical lobe structure may contribute to the brightness variations of an end-on lobe view.

\vspace{-0.3cm}

\section{Conclusions}
Our discovery of ORC J0102--2450 with ASKAP makes it the third odd radio circle with an elliptical galaxy at its geometrical centre. This is unlikely a coincidence and brings us a step closer to determining the ORC formation mechanisms. 
We discussed three scenarios, two of which have the central galaxy as its basis. These are (1) a relic lobe of a giant radio galaxy seen end-on or (2) a giant blast wave, possibly from a binary SMBH merger, resulting in a radio ring of $300-500$ kpc diameter. The third scenario considers radio galaxy and IGM interactions, involving the companion galaxy, that may be able to create the observed ring structure.

The discovery of further ORCs in the rapidly growing amount of wide-field radio continuum data from ASKAP and other telescopes will show if the above scenarios have any merit, contributing to exciting times in astronomy. Low-frequency LOFAR surveys at high-resolution (6$''$) will be of particular interest (see Shimwell et al. 2019), given the steep spectral index of known ORCs. Deep X-ray observations may also detect these energetic events as shown in the case of a giant relic radio galaxy by Tamhane et al. (2015).


\vspace{-0.3cm}

\section*{Acknowledgements}

We also thank the ASKAP team for their dedicated and continuing work on creating such a powerful survey telescope, together with a robust data processing pipeline and public archive. --- The Australian SKA Pathfinder is part of the Australia Telescope National Facility (ATNF) which is managed by CSIRO. Operation of ASKAP is funded by the Australian Government with support from the National Collaborative Research Infrastructure Strategy. ASKAP uses the resources of the Pawsey Supercomputing Centre. Establishment of ASKAP, the Murchison Radio-astronomy Observatory (MRO) and the Pawsey Supercomputing Centre are initiatives of the Australian Government, with support from the Government of Western Australia and the Science and Industry Endowment Fund. This paper includes archived data obtained through the CSIRO ASKAP Science Data Archive (CASDA). We acknowledge the Wajarri Yamatji as the traditional owners of the Observatory site.
--- This project used public archival data from the Dark Energy Survey (DES). Funding for the DES Projects has been provided by the U.S. Department of Energy, the U.S. National Science Foundation, the Ministry of Science and Education of Spain, the Science and Technology Facilities Council of the United Kingdom, the Higher Education Funding Council for England, the National Center for Supercomputing Applications at the University of Illinois at Urbana-Champaign, the Kavli Institute of Cosmological Physics at the University of Chicago, the Center for Cosmology and Astro-Particle Physics at the Ohio State University, the Mitchell Institute for Fundamental Physics and Astronomy at Texas A\&M University, Financiadora de Estudos e Projetos, Funda{\c c}{\~a}o Carlos Chagas Filho de Amparo {\`a} Pesquisa do Estado do Rio de Janeiro, Conselho Nacional de Desenvolvimento Cient{\'i}fico e Tecnol{\'o}gico and the Minist{\'e}rio da Ci{\^e}ncia, Tecnologia e Inova{\c c}{\~a}o, the Deutsche Forschungsgemeinschaft, and the Collaborating Institutions in the Dark Energy Survey. 
--- This publication makes use of data products from the Wide-field Infrared Survey Explorer, which is a joint project of the University of California, Los Angeles, and the Jet Propulsion Laboratory/California Institute of Technology, funded by the National Aeronautics and Space Administration. Partial support for LR comes from US NSF Grant AST 17-14205 to the University of Minnesota. --- Finally, We thank the referee and journal editors for their timely responses and constructive suggestions.  

\vspace{-0.3cm}

\section*{Data availability} 

The ASKAP data used in this article are available through the CSIRO ASKAP Science Data Archive (CASDA) under https://doi.org/10.25919/5e5d13e6bda0c. Additional data processing and analysis was conducted using the {\sc miriad} software\footnote{https://www.atnf.csiro.au/computing/software/miriad/} and the Karma visualisation\footnote{https://www.atnf.csiro.au/computing/software/karma/} packages. DES and WISE images were obtained through the Astro Data Lab\footnote{https://datalab.noirlab.edu/sia.php} and SkyView\footnote{https://skyview.gsfc.nasa.gov/current/cgi/query.pl} servers, respectively.










\bsp	
\label{lastpage}
\end{document}